\documentclass{llncs}
\usepackage{latexsym}
\usepackage{amsmath}
\usepackage{amssymb}
\usepackage{tikz}
\usetikzlibrary{arrows.meta}
\usepackage{graphicx}

\usepackage{epic}
\newcommand{\nat}{\mathbb{N}}
\newcommand{\desda}{\; \Longleftrightarrow \;}
\newcommand{\tl}{{\sf tail}}
\newcommand{\ar}{{\sf ar}}
\newcommand{\even}{{\sf even}}
\newcommand{\odd}{{\sf odd}}
\newcommand{\spir}{{\sf spir}}
\newcommand{\fib}{{\sf fib}}

\newcommand{\rep}{{\sf rep}}

\begin{document}
\title{Characterizing morphic sequences}

\author{Hans Zantema
} %\inst{1,2}}  \and Wieb Bosma\inst{2,3}}
\institute{Department of Computer Science, TU Eindhoven\footnote{From 2007 until his retirement in 2022 Zantema had an exchange position with Herman Geuvers, in which Zantema was employed for one day a week at Radboud University in Nijmegen and Geuvers for one day a week at TU in Eindhoven}
, P.O.\ Box 513,\\
5600 MB Eindhoven, the Netherlands, email: {\tt h.zantema@tue.nl}}

\authorrunning{H. Zantema}

%\keywords{morphic sequences, automata, rational terms, infinitary rewriting}

\maketitle

\begin{abstract}
Morphic sequences form a natural class of infinite sequences, extending the well-studied class of automatic sequences. Where automatic sequences are known to have several equivalent characterizations and the class of automatic sequences is known to have several closure properties, for the class of morphic sequences similar closure properties are known, but only limited equivalent characterizations. In this paper we extend the latter. We discuss a known characterization of morphic sequences based on automata and we give a characterization of morphic sequences by finiteness of a particular class of subsequences. Moreover, we relate morphic sequences to rationality of infinite terms and describe them by infinitary rewriting.
\end{abstract}

\section{Introduction}
The simplest class of infinite sequences over a finite alphabet are {\em periodic}: sequences of the shape $u^\infty$ for some finite non-empty word $u$.
The one-but-simplest are {\em ultimately periodic}: sequences of the shape $v u^\infty$ for some finite non-empty words $u,v$. But these are still boring in some sense. More interesting are well-structured sequences that are simple to define, but not being ultimately periodic.
One class of such sequences are {\em morphic sequences}, being the topic of this paper. As a basic example consider the morphism $f$ replacing 0 by the word $01$ and replacing 1 by the symbol 0. Then we obtain the following sequence of words:
\[ f(0) = 01,\]
\[ f^2(0) = f(f(0)) = f(01) = 010,\]
\[ f^3(0) = f(f^2(0)) = f(010) = 01001,\]
\[ f^4(0) = f(f^3(0)) = f(01001) = 01001010,\]
and so on. We observe that in this sequence of words for every word its predecessor is a prefix, so we can take the limit of the sequence of words, being the binary {\em Fibonacci sequence} $\fib$. This is a typical example of a (pure) morphic sequence. In general, a {\em pure morphic sequence} over a finite alphabet $\Gamma$ is of the shape $f^\infty(a)$ for a finite alphabet $\Gamma$ and
 $f : \Gamma \to \Gamma^+$, $a \in \Gamma$ and $f(a) = au$, $u \in \Gamma^+$. Here $f^\infty(a)$ is defined as the limit of $f^n(a)$, which is well-defined as for every $n$ the word $f^n(a) = a u f(u) \cdots f^{n-1}(u)$ is a prefix of $f^{n+1}(a) = a u f(u) \cdots f^{n}(u)$. So
\[ f^\infty(a) \; = \; a u f(u) f^2(u) f^3(u) \cdots.\]
If we have a (typically smaller) finite alphabet $\Sigma$, and a {\em coding} $\tau : \Gamma \to \Sigma$, then a {\em morphic sequence} over the alphabet $\Sigma$ is defined to be of the shape $\tau(\sigma)$. Clearly any pure morphic sequence is morphic by choosing $\Sigma = \Gamma$ and $\tau$ to be the identity. Basic properties of morphic sequences are extensively described in the books \cite{AS03,R14a}. A characterization of morphic sequences based on automata is given in \cite{R14b}.
Papers relating morphic sequences to rewriting and checking properties automatically include \cite{Z09,ZR10,ZE11}.

A typical example of a morphic sequence that is not pure morphic is
\[ \spir \; = \; 1101001000100001\cdots ,\]
consisting of infinitely many ones and for which the numbers of zeros in between two successive ones is $0,1,2,3,\ldots$, respectively.
Choosing $\Gamma = \{0,1,2\}$ and $f(2) = 21, f(1) = 01, f(0) = 0$, we obtain $f^2(2) = 2101$, $f^3(2) = 2101001$ and so on, yielding
\[ f^\infty(2) \; = \; 2101001000100001\cdots ,\]
for which indeed by choosing $\Sigma = \{0,1\}$, $\tau(0) = 0, \tau(1)= \tau(2) = 1$ we obtain $\spir = \tau(f^\infty(2))$, showing that
$\spir$ is morphic. But $\spir$ is not pure morphic. If it was then we have $\spir = f^\infty(1)$ for $f$ satisfying $f(1) = 1u$, $f(0) = v$, yielding contradictions for all cases: $u$ should contain a $0$, but no $1$ (otherwise $f^\infty(1)$ would contain a pattern $10^k1$ infinitely often for a fixed $k$), and $v$ should contain a $1$ (otherwise $f^\infty(1)$ would contain only a single $1$), but as $f^\infty(1)$ purely consists of $f(0) = v$ and $f(1) = 1u$ it cannot contain unbounded groups of zeros.

If you see this definition of morphic sequences for the first time, it looks quite ad hoc. However, there are several reasons to consider this class of morphic sequences as a natural class of sequences. One of them is the observation that the class of morphic sequences is closed under several kinds of operations. For instance, if the result of any morphism $g : \Sigma \to \Sigma^*$ applied on any morphic sequence over $\Sigma$ is an infinite sequence, then it is always again morphic. A more general result states that if applying a {\em finite state transducer} on a morphic sequence yields an infinite sequence, then it is always morphic too. Results like these are not easy, and can be found in \cite{AS03}, Corollary 7.7.5 and Theorem 7.9.1.

The class of morphic sequences is a generalization of the class of {\em $k$-automatic sequences}, being the main topic of \cite{AS03}.
These $k$-automatic sequences can be defined in several ways, but one way corresponds to our definition of morphic sequences with the extra requirement that the morphism $f : \Gamma \to \Gamma^+$ is {\em $k$-uniform}, that is, the length $|f(b)|$ of $f(b)$ is equal to $k$ for every $b \in \Gamma$.

As can be found in \cite{AS03}, this class of $k$-automatic sequences can be defined in several equivalent ways. One way is by means of an automaton (justifying the name `automatic'), another one is by finiteness of the $k$-kernel of the sequence, being a particular set of subsequences of the sequence. This has the flavor of a main well-known result from formal language theory, namely that the class of {\em regular languages} can be defined in several equivalent ways: DFAs, NFAs, regular expressions, right-linear grammars, and so on.

Similar characterizations for the class of morphic sequences are less known. A main goal of this paper is to give an overview of such characterizations. Apart from the closure properties, having several equivalent characterizations indicate that the class of morphic sequences is a natural class of sequences to consider, just like the class of regular languages is a natural class of languages to consider.

In \cite{R14b}, page 76, Theorem 2.24, an automaton-based characterization for morphic sequences is given, generalizing the automaton-based characterization of $k$-automatic sequences. For $k$-automatic sequences the corresponding type of automaton is a DFAO, a DFA with output, over the alphabet $\{0,1,\ldots,k-1\}$, in which the transition function $\delta : Q \times \Sigma \to Q$ is total, where $Q$ is the set of states and $\Sigma$ is the alphabet. In the generalization this transition function is partial. More precisely, when numbering the symbols from $\Sigma$ from 0 to $n-1$, then every state $q \in Q$ has an arity $0< \ar(q) < n$, and $\delta(q,i)$ is defined if and only if
$i < \ar(q)$. Surprisingly, this is exactly the same notion of a {\em mix-DFAO} as it is used in \cite{EGH13} to define and investigate {\em mix-automatic sequences}, being an extension of automatic sequences closed under zip operations. But the way in which words are entered to the automaton is different. One of the main results of \cite{EGH13} is that morphic sequences and mix-automatic sequences are incomparable.

Another characterization more in the flavor of finiteness of the kernel that we give is that we show how an infinite tree structure on the natural numbers defines for every natural number a subsequence of a given sequence, and we show that a sequence is morphic if and only if a tree structure exists such that the set of corresponding subsequences is finite.

Both sequences and these infinite trees can be seen as infinite terms: for sequences it is only over unary symbols, and for trees also symbols of arity $> 1$ appear. Infinite terms are called {\em rational} if they only contain finitely many distinct subterms. We reformulate our result on finitely many distinct subsequence in terms of rationality of terms, and we also describe morphic sequences as infinitary normal forms using infinitary rewriting.

Another more esthetic argument to consider morphic sequences is that they give rise to interesting {\em turtle graphics}, \cite{Z16}.
Having a sequence $\sigma$ over a finite alphabet $\Sigma$, one chooses an angle $\phi(b)$ for every $b \in \Sigma$. Next a picture is drawn in the following way: choose an actual angle that is initialized in some way, and next proceed as follows: for $i = 0,1,2,3,\ldots$ the actual angle is turned by $\phi(\sigma(i))$ and after every turn a unit segment is drawn in the direction of the actual angle. For every segment its end point is used a the starting point for the next segment. In this way infinitely many segments are drawn. The resulting figure is called a {\em turtle figure}.

In \cite{Z16} turtle figures are investigated for several morphic sequences, sometimes yielding finite pictures, that is, after a finite but typically very big number of steps only segments will be drawn that have been drawn before by which the picture will not change any more.
In other cases the turtle figure will be {\em fractal}.

To give the flavor of these figures, we show the turtle figure for the very simple pure morphic (and even 2-automatic) sequence $f^\infty(0) = 010001010100\cdots$ for $f(0) = 01$, $f(1) = 00$, sometimes called the {\em period doubling sequence}, and angles $\phi(0) = 140^o$ and $\phi(1) = -80^o$, so for every 0 the angle turns 140 degrees $= \frac{7 \pi}{9}$ to the left and for every 1 the angle turns 80 degrees $= \frac{4 \pi}{9}$ to the right, while after every turn a unit segment is drawn. The result is as follows:

\begin{center}
\includegraphics[width=120mm]{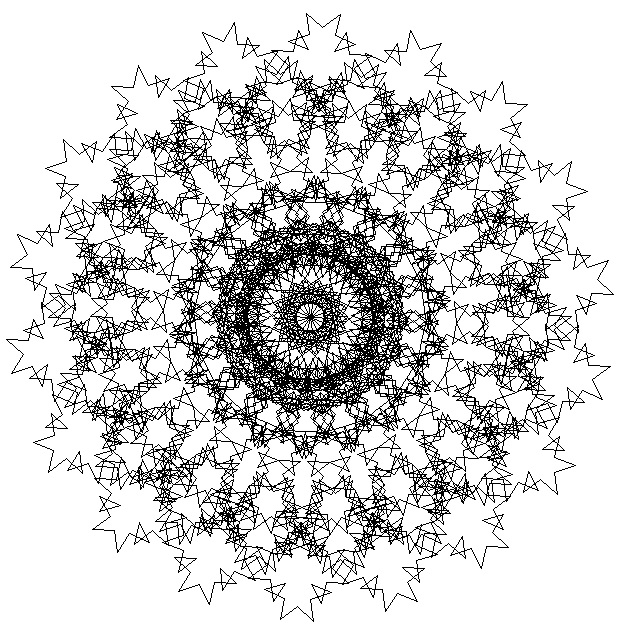}
\end{center}

A simple program of only a few lines that draws this particular turtle figure shows that after 6000 steps not yet the full figure is drawn, but in \cite{Z23,Z24} it is shown that after $9216 = 2^{10}\times 3^2$ steps, only segments will be drawn that were drawn before, so the full picture will be equal to the picture drawn by only this finite part, and that is the picture we show here. Other morphic sequences (many of which not being automatic) and other angles give rise to a wide range of remarkable turtle figures, many of which are shown in \cite{Z16,Z23,Z24}. Turtle figures can be made for all kinds of sequences, but morphic sequences show up a nice balance between the boring and very regular figures for ultimately periodic sequences on the one hand, and the complete chaos of random sequences on the other hand. The notation for the sequence $\spir$ we saw is motivated by the fact that it gives rise to spiral shaped turtle figures.

This paper is organized as follows. In Section \ref{secprel} we give some notation and preliminaries. In Section \ref{sectree} we show how the definition of a (pure) morphic sequence gives rise to a tree structure on the natural numbers. This tree structure is fully defined by a weakly monotone and surjective function $P : \nat^+ \to \nat$ that plays an important role in the rest of the paper.
The we give the various equivalent characterizations, starting in Section \ref{secaut} where we characterize morphic sequences by means of mix-DFAOs, a particular kind of automata. Next in Section \ref{secsubs} we show that morphic sequences are characterized by the property that a particular class of subsequences consist of only finitely many distinct sequences. In Section \ref{secinft} the same result is reformulated in terms of infinite terms. In Section \ref{secinfr} we present how morphic sequences can be characterized by infinitary rewriting. We conclude in Section \ref{secconcl}.

\section{Notation and preliminaries}
\label{secprel}

We write $\nat$ for the set of natural numbers (including $0$) and $\nat^+$ for the set of natural numbers $> 0$.
A function $f : \nat \to \nat$ is called {\em weakly monotone} if $f(n+1) \geq f(n)$ for all $n \in \nat$.

For a {\em word} $w \in \Sigma^*$ over an alphabet $\Sigma$ we write $|w|$ for the length of $w$. The set of non-empty words over $\Sigma$ is denoted by $\Sigma^+$. The empty word of length 0 is denoted by $\epsilon$.

A {\em sequence} $\sigma$ over an alphabet $\Sigma$ can be seen as a mapping $\sigma : \nat \to \Sigma$, so
\[ \sigma \; = \; \sigma(0) \sigma(1) \sigma(2) \cdots .\]
So by sequence we always mean an infinite sequence. In some texts they are called {\em streams} or {\em infinite words}.
The set of sequences over $\Sigma$ is denoted by $\Sigma^\infty$.

Morphisms $f : \Sigma \to \Sigma^+$ can also be applied on words and sequences, for instance, $f(\sigma)$ is the concatenation of the following words:
\[ f(\sigma) \; = \; f(\sigma(0))f(\sigma(1))f(\sigma(2)) \cdots.\]
Apart from applying morphisms another simple operation on sequences is the tail $\tl$ that simply removes the first element, so
\[ \tl(\sigma) \; = \; \sigma(1) \sigma(2) \sigma(3) \cdots.\]

From the introduction we recall:
\begin{definition}
A {\em pure morphic sequence} over a finite alphabet $\Gamma$ is of the shape $f^\infty(a)$ for a finite alphabet $\Gamma$ and
 $f : \Gamma \to \Gamma^+$, $a \in \Gamma$ and $f(a) = au$, $u \in \Gamma^+$.

A {\em morphic sequence} over a finite alphabet $\Sigma$ is of the shape $\tau(\sigma)$ for some coding $\tau : \Gamma \to \Sigma$ and some pure morphic sequence $\sigma$ over some finite alphabet $\Gamma$.
\end{definition}

Here $f^\infty(a)$ is defined as the limit of $f^n(a)$, which is well-defined as for every $n$ the word $f^n(a) = a u f(u) \cdots f^{n-1}(u)$ is a prefix of $f^{n+1}(a) = a u f(u) \cdots f^{n}(u)$. So
\[ f^\infty(a) \; = \; a u f(u) f^2(u) f^3(u) \cdots.\]
It is easy to see that it is the unique {\em fixed point} of $f$ starting in $a$, that is, $f(f^\infty(a)) = f^\infty(a)$.

In the above notation the coding $\tau : \Gamma \to \Sigma$ is lifted to $\tau : \Gamma^\infty \to \Sigma^\infty$ in which $\tau$ is applied on alle separate elements.

A slightly more relaxed definition of morphic sequence also allows $f(b) = \epsilon$ for some elements $b \in \Gamma$. In \cite{AS03}, Theorem 7.5.1, it is proved that this is equivalent to our definition, so we may and will always assume that $f(b) \in \Gamma^+$ for all $b \in \Gamma$.

A sequence over $\Sigma$ can also be seen as an {\em infinite term} over $\Sigma$ in which every symbol $b \in \Sigma$ has {\em arity} 1, notation $\ar(b) = 1$. More general, if every symbol $b \in \Sigma$ has arity $\ar(b) \geq 1$, an infinite term over $\Sigma$
 is defined by saying which symbol is on which position. Here a {\em position} $p \in \nat^*$
is a sequence of natural numbers.  In order to be a proper
term, some requirements have to be satisfied as indicated in
the following definition. Here we write $\bot$ for undefined, use $f$ for symbols in $\Sigma$
and every symbol $f$ has an arity $\ar(f) \in \nat^+$.

\begin{definition}
An infinite {\em term} over a signature $\Sigma$
is defined to be a map $t : \nat^* \to \Sigma \cup \{\bot\}$
such that
\begin{itemize}
\item the root $t(\epsilon)$ of the term $t$ is defined, so
$t(\epsilon) \in \Sigma$, and
\item for all $p \in \nat^*$ and all $i \in \nat$ we have
\[ t(pi) \in \Sigma \desda t(p) \in \Sigma \wedge 0 \leq i < \ar(t(p)). \]
\end{itemize}

A word $p \in \nat^*$ such that $t(p) \in \Sigma$ is called a {\em position} of $t$.

The {\em subterm} $t_p : \nat^* \to \Sigma \cup \{\bot\}$ at position $p \in \nat^*$ satisfying $t(p) \in \Sigma$ is defined by
$t_p(q) = t(pq)$ for all $q \in \nat^*$.

An infinite term is called {\em rational} if it contains only finitely many distinct subterms.
\end{definition}

When allowing {\em constants}, that is, symbols $f$ with $\ar(f) = 0$, this definition would also cover usual finite terms. As we only consider infinite terms with $\ar(f) > 0$ for all $f \in \Sigma$, we do not allow constants and all our terms are defined on infinitely many positions including $0^n$ for all $n \in \nat$. In many presentations positions are words over numbers from 1 to $\ar(f)$. Due to the relationship with number representations we prefer to number from 0 to $\ar(f)-1$, just like digits in decimal numbers are numbered from 0 to 9 and from 1 to 10.

A sequence $\sigma : \nat \to \Sigma$ now can be seen as an infinite term $t_\sigma$ over $\Sigma$ where every symbol has arity $1$, and any number $n \in \nat$ is identified with the position $0^n$, so $t_\sigma(n)(0^n) = \sigma(n)$ for all $n \in \nat$ and $t_\sigma(p) = \bot$ for all $p \in \nat^*$ not being of the shape $0^n$.

\section{Tree structures of morphic sequences}
\label{sectree}

In this section we show how a (pure) morphic sequence $f^\infty(a)$ gives rise to a tree structure on the natural numbers.
Mapping every natural to its position in the tree may be seen as an {\em enumeration system}, being the main topic of
\cite{R14a,R14b}.

A pure morphic sequence $f^\infty(a)$ is the unique fixed point of $f$ starting in $a$, and is of the shape
\[ au f(u) f^2(u) f^3(u) \cdots.\]
For every $i > 0$ the  element in $f^\infty(a)(i)$ is obtained as an element of $f(b)$ for some $b = f^\infty(a)(j)$ for $j < i$, so occurring earlier in the sequence. If we draw an arrow between every element in the sequence and the element from which it is obtained, we
get a tree structure, as is illustrated in the following example, where $\Gamma = \{0,1\}$, $f(0) = 01$, $f(1) = 0$. So the sequence $f^\infty(0)$ is the binary Fibonacci sequence $\fib$ as we saw in the introduction. The first few levels of the tree look as follows:

\begin{center}
	\begin{tikzpicture}
	
\node[circle,draw,inner sep=0pt,minimum width=5mm] (0) at (5,5) {0};
\node[circle,draw,inner sep=0pt,minimum width=5mm] (1) at (5,4) {1};
\node[circle,draw,inner sep=0pt,minimum width=5mm] (2) at (5,3) {0};
\node[circle,draw,inner sep=0pt,minimum width=5mm] (3) at (4,2) {0};
\node[circle,draw,inner sep=0pt,minimum width=5mm] (4) at (6,2) {1};
\node[circle,draw,inner sep=0pt,minimum width=5mm] (5) at (3,1) {0};
\node[circle,draw,inner sep=0pt,minimum width=5mm] (6) at (5,1) {1};
\node[circle,draw,inner sep=0pt,minimum width=5mm] (7) at (7,1) {0};
\node[circle,draw,inner sep=0pt,minimum width=5mm] (8) at (2,0) {0};
\node[circle,draw,inner sep=0pt,minimum width=5mm] (9) at (3.6,0) {1};
\node[circle,draw,inner sep=0pt,minimum width=5mm] (10) at (5,0) {0};
\node[circle,draw,inner sep=0pt,minimum width=5mm] (11) at (6.4,0) {0};
\node[circle,draw,inner sep=0pt,minimum width=5mm] (12) at (8,0) {1};
\node (a) at (0,5) {$a = 0$};
\node (u) at (0,4) {$u = 1$};
\node (fu) at (0,3) {$f(u) = 0$};
\node (f2u) at (0,2) {$f^2(u) = 01$};
\node (f3u) at (0,1) {$f^3(u) = 010$};
\node (f4u) at (0,0) {$f^4(u) = 01001$};
\node (0n) at (5.4,5) {0};
\node (1n) at (5.4,4) {1};
\node (2n) at (5.4,3) {2};
\node (3n) at (4.4,2) {3};
\node (4n) at (6.4,2) {4};
\node (5n) at (3.4,1) {5};
\node (6n) at (5.4,1) {6};
\node (7n) at (7.4,1) {7};
\node (8n) at (2.4,0) {8};
\node (9n) at (4,0) {9};
\node (10n) at (5.5,0) {10};
\node (11n) at (6.9,0) {11};
\node (12n) at (8.5,0) {12};

\draw[-] (0) -- (1);
\draw[-] (1) -- (2);
\draw[-] (2) -- (3);
\draw[-] (2) -- (4);
\draw[-] (3) -- (5);
\draw[-] (3) -- (6);
\draw[-] (4) -- (7);
\draw[-] (5) -- (8);
\draw[-] (5) -- (9);
\draw[-] (6) -- (10);
\draw[-] (7) -- (11);
\draw[-] (7) -- (12);

	\end{tikzpicture}
\end{center}

On the first level we have $a$, in this example being $0$, denoted inside the node, numbered by 0. On the second level we have $u$, in this example being $1$, again inside the node, numbered by 1. If $|u| > 1$ this level does not consist of a single node but of $|u|$ nodes.
On the next levels we have $f(u)$, $f^2(u)$, and so on. So the full pure morphic sequence $f^\infty(a) = auf(u) f^2(u) \cdots$ is obtained by concatenating all node labels in the order of the node numbers. This applies similarly for any morphic sequence, not only for this example.

The tree can be seen as an infinite term over the signature $\Gamma \cup \{a_0\}$, where the arity of a symbol $b \in \Gamma$ is $|f(b)|$ and the arity of $a_0$ is $|f(a)| -1$. Here $a_0$ only occurs at the root.

Fixing $f$ and $a$, we write $P : \nat^+ \to \nat$ for the function mapping $i$ to its parent in the tree. So $P(i) = j$ if $f^\infty(a)(i)$ is obtained from $f(b)$ for $b = f^\infty(a)(j)$. We write $R : \nat^+ \to \nat$ for the function mapping $i$ to the corresponding position in $f(b)$, being 0 for the first position and
$|f(b)| - 1$ for the last. For instance, the top right 1 numbered by 12 is $f^\infty(0)(12)$, obtained as the second element of $f(0)$ for $0$ being $f^\infty(0)(7)$, so $P(12) = 7$ and $R(12) = 1$.

These functions $P,R$ can be defined inductively as follows:
\begin{itemize}
\item $P(1) = 0,\; R(1) = 1$,
\item $P(n+1) = P(n)$ and $R(n+1) = R(n)+1$ if $R(n)+1 < |f(b)|$ for $b = f^\infty(a)(P(n))$,
\item $P(n+1) = P(n)+1$ and $R(n+1) = 0$ if $R(n)+1 = |f(b)|$ for $b = f^\infty(a)(P(n))$.
\end{itemize}
For $f$ being $k$-uniform, that is, $|f(b)| = k > 1$ for all $b \in \Gamma$, we have $P(i) = i \div k$ and $R(i) = i \mod k$ for all $i > 0$.

We collect some basic properties:
\begin{theorem}
\label{thmmorph}
For $f : \Gamma \to \Gamma^+$, $a \in \Gamma$ and $f(a) = au$, $u \in \Gamma^+$ and $P,R$ as defined above, we have
\begin{enumerate}
\item $P : \nat^+ \to \nat$ is weakly monotone and surjective, and $P(n) < n$ for every $n > 0$,
\item $f^\infty(a)(n) \; = \; f(b)(R(n))$ for $b = f^\infty(a)(P(n))$, for every $n > 0$,
\item $R(n) = \max \{ k | P(n-k) = P(n)\}$ for every $n > |u|$.
\item The infinite term representing the corresponding tree is rational.
\end{enumerate}
\end{theorem}
\begin{proof}
1: Using $P(n+1)$ is either $P(n)$ or $P(n+1)$ for all $n \in \nat$, we conclude weak monotonicity and prove $P(n) < n$ by induction on $n$. For surjectivity we use the same property, and the fact that for every $n$ there are only finitely many $k$ satisfying $P(k) = n$.

2 and 3: immediate.

4: Except for the root, two nodes with the same symbol from $\Gamma$ represent the same subtree, so rationality follows from finiteness of $\Gamma$. \qed
\end{proof}

For all cases, the tree structure can be fully obtained by only the function $P : \nat^+ \to \nat$: the nodes correspond to $\nat$, the root corresponds to 0, and for every $n > 0$ the parent of $n$ is $P(n)$.

For the representation $\tau(f^\infty(a))$ of a morphic sequence its tree structure is defined to be the tree defined by the function $P : \nat^+ \to \nat$ for the pure morphic sequence $f^\infty(a)$. Later on (in Section \ref{secdistrepr}) we will see that distinct representations of a morphic sequence may have distinct tree structures, and some distinct morphic sequences may be represented by the  same tree structure.

\section{Morphic sequences characterized by automata}
\label{secaut}

From \cite{AS03} and many other sources we recall the definition of a DFAO:

\begin{definition}
A {\em deterministic finite automaton with output} (DFAO) is a sixtuple $(Q,q_0,\Sigma,\delta,\Gamma,\lambda)$, where
\begin{itemize}
\item $Q$ is a finite set of {\em states},
\item $q_0 \in Q$ is the {\em initial state},
\item $\Sigma$ is a finite {\em alphabet},
\item $\delta : Q \times \Sigma \to Q$ is a {\em transition function},
\item $\Gamma$ is a finite {\em output alphabet},
\item $\lambda : Q \to \Gamma$ is the {\em output function}.
\end{itemize}
\end{definition}

From \cite{R14b} we recall the definition of ANS:

\begin{definition}
An {\em abstract numeration system} (ANS) is a threetuple $(L,A,<)$, where
\begin{itemize}
\item $A$ is a finite {\em alphabet},
\item $L$ is a regular language over $A$, and
\item $<$ is a total order on $A$.
\end{itemize}
An ANS $S = (L,A,<)$ defines the bijective representation function $\rep_S : \nat \to L$ numbering the elements of $L$ ina lexicographic way, more precisely, for all $i<j$ we have either $|\rep_S()i) < |\rep_S(j)|$, or $|\rep_S()i) = |\rep_S(j)|$ and $\rep_S(i) <^{\rm{lex}} \rep_S(j)$.
\end{definition}

Now we give the main theorem characterizing morphic sequences by automata, being Theorem 2.24 from \cite{R14b}; for the proof we refer to \cite{R14b}.

\begin{theorem}
\label{thmsaut}
A sequence $\sigma$ over an alphabet $\Gamma$ is morphic if and only if a DFAO $(Q,q_0,\Sigma,\delta,\Gamma,\lambda)$ and an ANS $S = (L,A,<)$ exists such that such that for every $i \in \nat$ one has $\sigma(i) = \lambda(\delta(q_0,\rep_S(i)))$.
\end{theorem}

Next we present a variant of Theorem \ref{thmsaut} we independently found ourselves, not referring to ANS, but instead finetuning the notion of DFAO to mix-DFAO. The difference with a standard DFAO is that in a mix-DFAO the number of outgoing arrows from a state is not everywhere the same but may depend on the state.

\begin{definition}
A {\em mix deterministic finite automaton with output} (mix-DFAO) is a sixtupel $(Q,\ar,\delta,q_0,\Sigma,\lambda)$, where
\begin{itemize}
\item $Q$ is a finite set of {\em states},
\item $\ar : Q \to \nat^+$ is the {\em arity} function,
\item $\delta : Q \times \nat \to Q$ is a partial {\em transition function} for which $\delta(q,a)$ is defined if and only if $0 \leq a < \ar(q)$,
\item $q_0 \in Q$ is the {\em initial state},
\item $\Sigma$ is a finite {\em alphabet}, and
\item $\lambda : Q \to \Sigma$ is the {\em output function}.
\end{itemize}
\end{definition}

In \cite{EGH13} mix-DFAO's were used to define and investigate {\em mix-automatic sequences}, being an extension of automatic sequences closed under zip operations. Here the $i$th element of the sequence is obtained by feeding the interpretation of $i$ in some dynamic radix enumeration system to the automaton and taking the output of the resulting state. So the key idea is to represent $i$ by a word $w$, and feed $w$ to the automaton. 

Surprisingly, we will use exactly the same notion of mix-DFAO to characterize morphic sequences. Given such a mix-DFAO $M$ we will define a sequence $\sigma_M$ over $\Sigma$ that will be morphic. We do this by defining $\sigma_M(i) \in \Sigma$ for every $i \in \nat$. As such an element in $\Sigma$ may be defined as $\lambda(q)$ for some $q \in Q$, we do this by mapping natural numbers to states, using the transition function. This idea is the same as for mix-automatic sequences. The difference is that we use a different representation of the number $i$ by a word. While for mix-automatic sequences the dynamic radix interpretation is taken from right to left, starting at the least significant digit, we will do it the other way around, so starting at the most significant digit.
For $k$-automatic sequences these correspond to the two ways to define such a sequence by a DFAO: a natural number $i$ is either mapped to $\delta(q_0,w_i)$ or to $\delta(q_0,w^R_i)$, where $w_i$ is the $k$-ary notation of $i$, and every state has arity $k$, and $R$ stands for reversing the word. In our morphic setting the representation is slightly more complicated as different states may have different arities.

Let $n_M = \max \{\ar(q) \mid q \in Q \}$ and $\Delta = \{ 0,1,\ldots, n_M - 1\}$. As $\delta(q,a)$ is undefined for every $a \not\in \Delta$, we may consider $\delta$ as a partial function $\delta : Q \times \Delta \to Q$. As usual in automata we extend $\delta$ allowing words in the second argument rather than single symbols. Due to the partial character of $\delta$, the result of $\delta$ on a word may be undefined. We define $L_M$ to be the set of words $w$ over $\Delta$ for which $\delta(q_0,w) \in Q$ is defined, more precisely, it is defined as the smallest prefix closed language over $\Delta$ satisfying $\epsilon \in L_M$ and $w a \in L_M$ for every $w \in L_M$ and $0 \leq a < \ar(\delta(q_0,w))$.

We define the enumeration function $\phi_M : \nat \to L_M$ inductively as follows: $\phi_M(0) = \epsilon$ and $\phi_M(1) = 1$. For the inductive part we have the invariant that if $\phi_M(n) = w a$ for $w \in L_M$ and $a \in \Gamma$, then there exists exactly one $n' < n$ satisfying $\phi_M(n') = w$. Indeed for $n=1, n' =0$ this holds for $w = \epsilon$. For defining $\phi_M(n+1)$ for $n > 0$ write $\phi_M(n) = w a$. In case $a+1 < \ar(\delta(q_0,w)$ then we define
$\phi_M(n+1) = w (a+1)$, otherwise we define $\phi_M(n+1) = \phi_M(n' + 1) 0$ for $n' < n$ satisfying $\phi_M(n') = w$.

Note that for the $k$-automatic case where $\ar(q) = k$ for all $q \in Q$, $\phi_M(n)$ coincides with the $k$-ary notation of $n$.

For a mix-DFAO $M = (Q,\ar,\delta,q_0,\Sigma,\lambda)$ the sequence $\sigma_M$ over $\Sigma$ is defined by
\[ \sigma_M(i) \; = \; \lambda(\delta(q_0,\phi_M(i))) \]
for all $i \in \nat$. Now we are ready to formulate our characterization of morphic sequences by mix-DFAOs.

\begin{theorem}
\label{thmaut}
A sequence $\sigma$ over an alphabet $\Sigma$ is morphic if and only if a mix-DFAO $M = (Q,\ar,\delta,q_0,\Sigma,\lambda)$ exists such that $\sigma = \sigma_M$.
\end{theorem}

\begin{proof}
Assume that $\sigma$ is morphic over $\Sigma$. Then  $\sigma = \tau(f^\infty(a))$ for a finite alphabet $\Gamma$ and
$\tau : \Gamma \to \Sigma$, $f : \Gamma \to \Gamma^+$, $a \in Q$ and $f(a) = au$, $u \in Q+$. We define the mix-DFAO $M = (Q,\ar,\delta,q_0,\Sigma,\lambda)$ as follows:
\begin{itemize}
\item $Q = \Gamma$,
\item $\ar(q) = |f(q)|$ for all $q \in Q$,
\item $\delta(q,i) = f(q)(i)$, so the $(i+1)$th element of $f(q)$ for $q \in Q$, $0 \leq i < \ar(q)$,
\item $q_0 = a$,
\item $\lambda(q) = \tau(q)$ for $q \in Q$.
\end{itemize}
Now we claim that $\sigma = \tau(f^\infty(a)) = \sigma_M$. This follows since for every $n$ the following holds:

(1) $\delta(q_0,\phi_M(n)) = f^\infty(a)(n)$ and (2) $\phi_M(n) = wi$ then $\phi_M(P(n)) = w$ and $i = R(n)$,
where $P,R$ are the functions as defined for $f^\infty(a)$.

We prove this by induction on $n$. For $n=0$ (1) holds since $\delta(q_0,\phi_M(0)) = \delta(q_0, \epsilon) = q_0 = a = f^\infty(a)(0)$, and (2) holds since $\phi_M(n) = \epsilon$ is not of thew shape $wi$. For $n>0$ (2) is obtained by the definition of $\phi_M$ and the induction hypothesis, and for (1) the key observation is that for $\phi_M(n) = w i$ for $w \in L_M$, $0 \leq i < \ar(\delta(q_0),w)$, the unique $n' < n$ satisfying $\phi_M(n') = w$ coincides with $P(n)$, and $i$ coincides with $R(n)$.

%$\delta(q_0,\phi_M(i)) = f^\infty(a)(i)$ for all $i \in \nat$, that we prove by induction on $i$. For $i=0$ this holds since $\delta(q_0,\phi_M(0)) = \delta(q_0, \epsilon) = q_0 = a = f^\infty(a)(0)$. For $i>0$ the key observation is that for $\phi_M(n) = w i$ for $w \in L_M$, $0 \leq i < \ar(\delta(q_0),w)$, the unique $n' < n$ satisfying $\phi_M(n') = w$ coincides with $P(n)$, and $i$ coincides with $R(n)$. This follows from the following claim that is easily proved by induction on $n$: if $\phi_M(n) = wi$ then $\phi_M(P(n)) = w$ and $i = R(n)$.

As a consequence, we obtain $\delta(q_0, \phi_M(n)) = \delta(\delta(q_0,\phi_M(P(n))),R(n))$ for every $n>0$. Using this, we obtain for $i>0$:
\[\begin{array}{rcll}
\delta(q_0,\phi_M(i)) & = & \delta(\delta(q_0,\phi_M(P(i))),R(i)) & \\
& = & \delta(f^\infty(a)(P(i)),R(i)) & \mbox{by the induction hypothesis, $P(i) < i$} \\
& = & f(f^\infty(a)(P(i)))(R(i)) & \mbox{definition of $\delta$} \\
& = & f^\infty(a)(i) & \mbox{Theorem \ref{thmmorph}, part 2. } \end{array} \]
This proves one direction of the theorem.

For the other direction assume that $M = (Q,\ar,\delta,q_0,\Sigma,\lambda)$ is a mix-DFAO. Then we define $\Gamma = Q$, $a = q_0$, $f(q) = \delta(q,0) \delta(q,1) \cdots \delta(q,\ar(q)-1)$, and $\tau(q) = \lambda(q)$ for $q \in Q$. This defines the morphic sequence $\sigma = \tau(f^\infty(a)$, and we obtain $\sigma_M = \sigma$ by the same argument as above. \qed

\end{proof}

Note that $L_M$ is closely related to the set of words $p \in \nat^*$ for which $t(p) \neq \bot$ in the definition of infinite terms; the only difference is in the first symbol: non-empty words in $L_M$ always start in $b >0$, where in the position notation this is replaced by $b-1$.

Using the standard automaton notation where $\delta(q,i) = r$ is denoted by an arrow from node $q$ to node $r$ labeled by $i$, the start state $q_0$ is denoted by an incoming arrow, and the output $\lambda(q)$ of a state $q$ is denoted inside the node $q$, a DFAO for the binary Fibonacci sequence $f^\infty(0)$ for $f(0) = 01, f(1) = 0$ reads as follows:

\begin{center}
	\begin{tikzpicture}
	
\node[circle,draw,inner sep=0pt,minimum width=5mm] (0) at (1,0) {0};
\node[circle,draw,inner sep=0pt,minimum width=5mm] (1) at (4,0) {1};
\node (s) at (0,0) {};
\draw[->] (s) -- (0);
\draw[->] (0) -- node[below,inner sep=4pt] {$1$} (1);
\draw[->] (1) edge[out=135,in=45,looseness=1] node[above,inner sep=4pt] {$0$} (0);
\draw[->] (0) edge[out=225,in=315,looseness=8]  node[below,inner sep=4pt] {$0$} (0);
	\end{tikzpicture}
\end{center}

The tree we earlier gave for this sequence can be obtained by unwinding this automaton, starting from $q_0$ and ignoring the self-loop of $q_0$ at the first step. This observation holds for morphic sequences in general.

\section{Morphic sequences characterized by subsequences}
\label{secsubs}

As is presented in \cite{AS03}, $k$-automatic sequences can be characterized by finiteness of the $k$-kernel of the sequence. For $k=2$ this means that by applying the operations $\even$ and $\odd$ on the sequence any number of times in any order, only finitely many distinct sequences are obtained. In this section we show that the much more general class of morphic sequences can also be characterized by finiteness of some class of subsequences.

In Section \ref{sectree} we saw that the tree structure of a morphic sequence $\tau(f^\infty(a))$ is fully described by a surjective weakly monotone function $P : \nat^+ \to \nat$. Now we will use such a function $P$ to define subsequences corresponding to subtrees.

As a first observation we state the following lemma:

\begin{lemma}
Let $P : \nat^+ \to \nat$ be weakly monotone and surjective. Then for every $n \in \nat^+$ there exists $k > 0$ such that $P^i(n) > 0 $ for all $i < k$ and $P^k(n) = 0$.
\end{lemma}
\begin{proof}
We apply induction on $n$. For $n=1$ it holds since $P(1)=0$, which holds since $P(1) > 0$ would contradict surjectivity.

For $n> 1$ we have $P(n) < n$ as we already observed in Theorem \ref{thmmorph}, and the result follows from the induction hypothesis on $P(n)$. \qed
\end{proof}

A weakly monotone and surjective function $P : \nat^+ \to \nat$ we shortly call a {\em tree function}, as it defines a tree structure on $\nat$ in which $0$ is the root and $i$ is the parent of $j$ in the tree if and only if $P(j) = i$. For our purpose we need an extra requirement, namely that the tree is {\em rational}, that is, contains only finitely many distinct subtrees. A {\em rational tree function} is defined to be a tree function $P : \nat^+ \to \nat$ for which the corresponding tree is rational.

\begin{lemma}
\label{lemrat}
Let $P : \nat^+ \to \nat$ be the tree function corresponding to a pure morphic function $f^\infty(a)$ over a finite alphabet $\Sigma$ as described in Section \ref{sectree}. Then $P$ is a rational tree function.
\end{lemma}
\begin{proof}
Let $p,q$ be two nodes of the corresponding tree, both distinct from the root. Then if $p$ and $q$ are labeled by he same symbol from $\Sigma$, then by construction the subtrees having $p$ and $q$ as their roots are equal. Now the lemma follows from finiteness of $\Sigma$. \qed
\end{proof}

For instance, for the tree we presented for the binary Fibonacci sequence $\fib$ in Section \ref{sectree} there are three distinct subtrees: the full tree having node 0 as its root, and the two subtrees having nodes 1 and 2 as its root, respectively. Every other subtree of which the root is labeled by 0 is equal to subtree having 2 as its root, and every other subtree of which the root is labeled by 1 is equal to subtree having 1 as its root.

Let $P : \nat^+ \to \nat$ be a tree function. For $n \in \nat$ we define
\[S_P(n) = \{ m \in \nat \mid \exists k \in \nat : P^k(m) = n\}.\]
Note that for every $n \in \nat$ the set $S_P(n) \subseteq \nat$ is infinite.

For any infinite set $S \subseteq \nat$ we can uniquely write $S = \{s_0,s_1,s_2,\ldots\}$ with $s_0 < s_1 < s_2 < \cdots$. For a sequence $\sigma$ and an infinite set $S \subseteq \nat$ we define $\sigma^S$ being the subsequence
\[ \sigma^S \; = \; \sigma_{s_0}\sigma_{s_1}\sigma_{s_2} \cdots,\]
so $\sigma^S(i) = \sigma(s_i)$ for all $i \in \nat$.

\begin{theorem}
\label{thmsubseq}
A sequence $\sigma$ over $\Sigma$ is morphic if and only if a rational tree function $P : \nat^+ \to \nat$ exists such that the set
\[ \{ \sigma^{S_P(n)} \mid n \in \nat \} \]
of subsequences of $\sigma$ is finite.
\end{theorem}

\begin{proof}
First assume that $\sigma = \tau(f^\infty(a))$ is morphic. Then according to Theorem \ref{thmaut} there is
a mix-DFAO $M = (Q,\ar,\delta,q_0,\Sigma,\lambda)$ such that $\sigma = \sigma_M$. Let $P$ be the weakly monotone surjective function $P : \nat^+ \to \nat$ corresponding to $f,a$. According to Lemma \ref{lemrat} the function $P$ is a rational tree function. The key observation now is that for every $n$ we have $S_P(n) = \{ k \mid \phi_M(n) \mbox{ is a prefix of } \phi_M(k) \}$. This follows from the observation we made in the proof of Theorem \ref{thmaut}: for $n>0$ we have $\phi_M(n) = wi$ for $w = \phi_M(P(n))$ and $i = R(n)$. Now for every $n$ the sequence $\sigma^{S_P(n)}$ starts in $\sigma(n) = \lambda(\delta(q_0,\phi_M(n)))$ and continues by $\sigma(m) = \lambda(\delta(q_0,\phi_M(m)))$ for all $m$ for which
$\phi_M(m)$ has $\phi_M(n)$ as a prefix in a breadth-first manner. As this sequence observed in this way only depends on $\delta(q_0,\phi_M(n))$, we obtain $\sigma^{S_P(n)} = \sigma^{S_P(n)}$ if $\delta(q_0,\phi_M(n))= \delta(q_0,\phi_M(n))$. As $Q$ is finite, we conclude that $\{ \sigma^{S_P(n)} \mid n \in \nat \}$ is finite, proving one direction of the theorem.

For the converse assume that $\{ \sigma^{S_P(n)} \mid n \in \nat \}$ is finite for some rational tree function $P : \nat^+ \to \nat$ and some sequence $\sigma$ over $\Sigma$. Divide the elements of $\sigma$ in a breadth-first manner over the tree defined by $P$, that is, the root of the tree is labeled by $\sigma(0)$, the leftmost node $n$ with $P(n) = 0$ is labeled by $\sigma(1)$, and so. Then for every $n$ the node corresponding to $n$ is the root of the tree corresponding to $S_P(n)$, labeled by $\sigma^{S_P(n)}(0) = \sigma(n)$. Hence for every $n$ the value $\sigma(n)$ is obtained by following the path to the node numbered by $n$ in the tree. Apart from finiteness, the tree can now seen as a mix-DFAO $M$ such that $\sigma_M(n) = \sigma(n)$ for every $n \in \nat$. It remains to show that this infinite mix-DFAO can be collapsed to a real mix-DFAO, so being finite, having the same property.

We define $n,n'\in \nat$ to be equivalent, notation $n \sim n'$, if $\sigma^{S_P(n)} = \sigma^{S_P(n')}$ and the subtrees having roots $n,n'$ are equal. Now $\sim$ is an equivalence relation. Since $P$ is rational and $\{ \sigma^{S_P(n)} \mid n \in \nat \}$ is finite, the equivalence relation $\sim$ has finitely many equivalence classes. The observation now is that all numbers in the same equivalence class can be joined to a single state. Choose $V \subseteq \nat$ to be a set of representants of these equivalence classes, so a minimal finite set such that for every $n \in \nat$ there exists $n' \in V$ such that $n \sim n'$. Then the resulting mix-DFAO can be described as follows. Let $q_0 \not\in V$. The mix-DFAO $M = (Q,\ar,\delta,q_0,\Sigma,\lambda)$ is defined as follows:
\begin{itemize}
\item $Q = V \cup \{ q_0\}$,
\item $\ar(q) = | \{n \in \nat \mid P(n) = q \} |$ for $q \in V$, $\ar(q_0) = \ar(0)$,
\item $\delta(q_0,0) = q_0$, $\delta(q_0,i) = q$ for $q \in V$ for which $\sigma^{S_P(i)} = \sigma^{S_P(q)}$, for $i = 1,\ldots,\ar(q_0) - 1$,
\item $\delta(q,i) = r$ for $q,r \in V$, $i = 0,\ldots,\ar(q) - 1$, $\sigma^{S_P(x)} = \sigma^{S_P(r)}$, for $x = y+i$ where $y$ is the smallest value with $P(y) = q$,
\item $\lambda(q) = \sigma(q)$ for $q \in V$, $\lambda(q_0) = \sigma(0)$.
\end{itemize}
It is still the case that $\sigma_M(n) = \sigma(n)$ for all $n \in \nat$, proving that $\sigma$ is morphic by Theorem \ref{thmaut}. \qed
\end{proof}

For instance, for the binary Fibonacci sequence $\fib$ the set $\{ \sigma^{S_P(n)} \mid n \in \nat \}$ consists of three sequences:
$\sigma^{S_P(0)} = \fib$, $\sigma^{S_P(1)} = \tl(\fib)$ and  $\sigma^{S_P(2)} = \tl(\tl(\fib))$, corresponding to the three subtrees of the the rational tree we already observed. For more complicated examples of morphic sequences $\sigma$ typically $\{ \sigma^{S_P(n)} \mid n \in \nat \}$ also contains sequences that cannot be obtained from $\sigma$ by only applying the tail function $\tl$.

It is a natural question whether the requirement for the tree function $P$ being rational is essential in Theorem \ref{thmsubseq}.
It is, as is shown by the following construction. For any sequence natural numbers $n_1,n_2,n_3,\ldots$ of natural numbers satisfying $1\leq n_1 \leq n_2 \leq n_3 \leq \cdots$ we can make a tree $T$ for which the number of nodes with distance $i$ to the root is exactly $n_i$ for all $i \geq 1$. Next we consider the following tree:

\begin{center}
	\begin{tikzpicture}
	
\node[circle,draw,inner sep=0pt,minimum width=5mm] (0) at (5,0) {1};
\node[circle,draw,inner sep=0pt,minimum width=5mm] (1) at (4,1) {1};
\node[circle,draw,inner sep=0pt,minimum width=5mm] (2) at (3,2) {1};
\node[circle,draw,inner sep=0pt,minimum width=5mm] (3) at (2,3) {1};
\node (T1) at (6,1) {$T$};
\node (T2) at (5,2) {$T$};
\node (T3) at (4,3) {$T$};
\node (T4) at (3,4) {$\cdots$};
\node (4) at (1,4) {$\cdots$};

\draw[-] (0) -- (1);
\draw[-] (1) -- (2);
\draw[-] (2) -- (3);
\draw[-] (0) -- (T1);
\draw[-] (1) -- (T2);
\draw[-] (2) -- (T3);
\draw[-] (3) -- (4);
\draw[-] (3) -- (T4);

	\end{tikzpicture}
\end{center}

Here the indicated nodes are labeled by 1, and all nodes in all copies of $T$ are labeled by 0. Let $\sigma$ be the sequence and $P$ the tree function corresponding to this tree. Then $\{ \sigma^{S_P(n)} \mid n \in \nat \}$ consists only of two sequences: the sequence $\sigma$ described by any of all equal subtrees having any node labeled by $1$ at its root, and the sequence purely consisting of zeros having any node labeled by $0$ at its root. The sequence $\sigma$ contains infinitely many ones, and the numbers of zeros between them are $0, 1, 1 + n_1, 1+n_1+n_2, 1+n_1+n_2+n_3, \ldots$, respectively. But there are only countably many morphic sequences, and the uncountably many choices for $1\leq n_1 \leq n_2 \leq n_3 \leq \cdots$ yield uncountably many distinct sequences $\sigma$, so not all of these are morphic, showing that Theorem \ref{thmsubseq} does not hold if the requirement for $P$ being a rational tree function is weakened to only being any tree function.

We started this section by comparing our new construction of subsequences by the $k$-kernel of which finiteness is equivalent to being $k$-automatic. We want to stress that in case of the rational tree function $P : \nat^+ \to \nat$ corresponds to $k$-automatic sequences, so $P(n) = n \div k$, then the set $\{ \sigma^{S_P(n)} \mid n \in \nat \}$ does not coincide with the $k$-kernel. More precisely, the $k$-kernel consists of the subsequences $\sigma^S$ of $\sigma$ where $S$ runs over all numbers of which the $k$-ary notation ends with some fixed word $w$, while our subsequences $\sigma^{S_P(n)}$ run over the subsequences $\sigma^S$ of $\sigma$ where $S$ runs over all numbers of which the $k$-ary notation starts with some fixed word $w$.

\section{Morphic sequences characterized by rational infinite terms}
\label{secinft}

In Theorem \ref{thmsubseq} the subsequences $\sigma^{S_P(n)}$ are described by subtrees. In this section we reformulate this theorem by describing these subtrees as subterms of infinite terms. The idea is to represent the tree represented by the rational tree function $P : \nat^+ \to \nat$ by a rational term over the signature consisting of all finitely many subtrees, in which the arity corresponds to the number of children of the tree. Next the nodes of this tree are numbered in a breadth-first way, and for a sequence $\sigma$ every node with number $i$ is labeled by $\sigma(i)$. Now Theorem \ref{thmsubseq} essentially states that $\sigma$ is morphic if and only if this labeled tree is rational.

More precisely, we start by a rational tree function $P : \nat^+ \to \nat$. Let $\Delta$ consist of the finitely many subtrees, and the arity of $t \in \Delta$ is its number of children. Now the tree coincides with an infinite term $t_P$ over $\Delta$, described by a function $t_P : \nat^* \to \Delta \cup \{\bot\}$. We define a mix-DFAO $M$ in which the set of states is $\Delta$, the initial state is the full tree described by the root, the arity of every state is the number of children of the root of the corresponding subtree, and $\delta(q,i)$ is the $(i+1)$th child of the root of the corresponding subtree. The output function $\lambda$ does not play a role here. Next we define the function $\phi_M : \nat \to L_M$
as in Section \ref{secaut}, where $L_M$ is the language of words $w$ for which $\delta(q_0,w)$ is defined. Note that $L_M$ coincides exactly with the set of positions of the infinite term $t_P$. As $\phi_M : \nat \to L_M$ is a bijection, we can also consider its inverse
$\phi_M^{-1} : L_M \to \nat$, mapping every position to its number.

For any sequence $\sigma$ over a finite alphabet $\Sigma$, we extend  the signature $\Delta$ to $\Delta \times \Sigma$, where for every $(t,a) \in \Delta \times \Sigma$ the arity is $\ar(t)$. We consider the infinite term over this extended signature by labeling every node in the term $t_P$ by $\sigma(i)$, where $i$ is the number of the node.
More precisely, the term $t_P : \nat^* \to \Delta \cup \{\bot\}$ is extended to $t_{P,\sigma} : \nat^* \to \Delta  \times \Sigma \cup \{\bot\}$ defined by $t_{P,\sigma}(p) \; = \; (t_P(p),\sigma(\phi_M^{-1}(p)))$
for every $p \in \nat^*$.

Now the reformulation of Theorem \ref{thmsubseq} reads as follows:

\begin{theorem}
\label{thmterm}
A sequence $\sigma$ over the alphabet $\Sigma$ is morphic if and only if a rational tree function $P : \nat^+ \to \nat$ exists such that
the infinite term $t_{P,\sigma}$ is rational.
\end{theorem}

\section{Morphic sequences characterized by infinitary rewriting}
\label{secinfr}

In this section we describe how any morphic sequence can be obtained by infinitary rewriting. For basics on rewriting we refer to \cite{T03}. Infinitary rewriting extends rewriting to infinite terms and infinite reductions. Infinite terms were already introduced in our preliminaries. An infinite reduction is said to {\em converge} if for every $n$ it holds that after a finite number of steps no reductions take place any more on level $\leq n$. In that case the limit of the reduction is well-defined as an infinite term, and that is defined to be the {\em infinitary normal form} of the reduction.  For more basics on infinitary rewriting and such reductions to infinitary normal forms we refer to the chapter \cite{KV03} on infinitary rewriting in \cite{T03} and \cite{Z08}.

So the goal of this section is to describe any morphic sequence to be the infinitary normal form of a particular start term with respect to a particular rewrite system. The rewrite system is designed in such a way that it follows the definition of morphic sequence quite directly.

For a word $w = w_0 w_1 \cdots w_{k-1}$ of unary symbols $w_0,w_1, \ldots, w_{k-1}$ and a term $t$ we use $w(t)$ as an abbreviation for $w_0( w_1( \cdots (w_{k-1}(t) \cdots )))$.

\begin{theorem}
For a morphic sequence $\sigma = \tau(f^\infty(a))$ over $\Sigma$ for $\tau : \Gamma \to \Sigma$, $f : \Gamma \to \Gamma^+$, $f(a) = au$, $a \in \Gamma$, $u \in \Gamma^+$, we define the signature $\Delta = \Sigma \cup \Gamma \cup \overline{\Gamma} \cup \{S,E\}$, where $\overline{\Gamma} = \{ \overline{b} \mid b \in \Gamma \}$, the symbol $E$ is a constant marking the end, and all other symbols are unary.
The rewrite system $R_\sigma$ consists of the rules
\[ \begin{array}{rcll}
S(b(x)) & \to & \tau(b)(S(\overline{b}(x))) & \mbox{ for all $b \in \Gamma$,}\\
\overline{b}(c(x)) & \to & c(\overline{b}(x)) & \mbox{ for all $b,c \in \Gamma$,}\\
\overline{b}(E) & \to & f(b)(E) & \mbox{ for all $b \in \Gamma$}\end{array}\]

Then $\sigma$ is the unique infinitary normal form of $\tau(a)(S(u(E)))$.
\end{theorem}
\begin{proof}
For any word $w \in \Gamma^+$ and $b \in \Gamma$ we obtain $\overline{b}(w(E)) \to_{R_\sigma}^+ w(\overline{b}(E))$ by using the second type of rules. For any word $w \in \Gamma^+$ and $b \in \Gamma$ combining this by using the other types of rules yields
\[ S(b(w(E))) \to_{R_\sigma} \tau(b)(S(\overline{b}w(E)))  \to_{R_\sigma}^+ \tau(b)(S(w(\overline{b}(E)))) \to_{R_\sigma}
\tau(b)(S(w f(b)(E))).\]
Repeating this $|w|$ times yields $Sw(E) \to_{R_\sigma}^+ \tau(w) S f(w)(E)$ for any word $w \in \Gamma^+$. Hence
\[ \tau(a)(S(u(E))) \to_{R_\sigma}^+  \tau(au) S f(u) (E) \to_{R_\sigma}^+  \tau(au f(u)) S f^2(u) (E) \to_{R_\sigma}^+ \cdots \]
\[\to_{R_\sigma}^+ \tau(au f(u)\cdots f^{n-1}(u)) S f^n(u) (E) = \tau(f^n(a)) S f^n(u)(E)\]
for all $n > 0$. Due to the shape of the rules, in further rewriting $\tau(f^n(a)) S f^n(u)(E)$ the initial part $\tau(f^n(a))$ will not change any more, hence by repeating this for increasing $n$ converges to $\tau(f^\infty(a)) = \sigma$.

As $R_\sigma$ is orthogonal, no other infinitary normal form of $\tau(a)(S(u(E)))$ exists. \qed
\end{proof}

\section{Distinct representations for the same morphic sequence}
\label{secdistrepr}

The representation $\tau(f^\infty(a))$ of a morphic sequence is far from unique. A simple observation is that if $f$ is replaced by $f^n$ for some $n > 1$, we get the same sequence, so $\tau((f^n)\infty(a)) = \tau(f^\infty(a))$. The corresponding rational tree function is closely related: the function $P$ for $\tau(f^\infty(a))$ is just replaced by $P^n$. But for many morphic sequences there are several ways to represent them, and the relation among them is not clear. As an example we consider $\tl(fib) = 1001010\cdots$, that is, our well-known binary Fibonacci sequence from which the first element is removed. As is described in \cite{AS03}, Theorem 7.6.1,  the tail (there called {\em left shift} of any morphic sequence $\tau(f^\infty(a))$ for $f(a) = au$ is described as the morphic sequence $\tau(f^\infty(c))$ in which $c$ is a fresh symbol and $f$, $\tau$ are extended for $c$ by $f(c) = c y f(b)$ and $\tau(c) = \tau(b)$, where $b$ is the first symbol $u$ and $y$ is the rest of $u$, so $u = by$. For $\fib$ this gives
\[ \tl(\fib) \; = \; \tau(f^\infty(c))\]
for $f(0) = 01, f(1) = 0, f(c) = c0$ and $\tau(0) = 0, \tau(1) = \tau(c) = 1$. Observe that this representation for $\tl(fib)$ gives rise to another tree structure than the tree structure of our representation of $\fib$, for instance, the single child of its root has two children, while in our representation of $\fib$ the single child of its root has only one child.

Next we give another representation of $\tl(\fib)$ having exactly the same tree structure as our representation of $\fib$.
\begin{theorem}
\label{thmtlfib}
For $g, \rho$ defined by $g(0) = 2$, $g(1) = g(2) = 10$, $\rho(0) = \rho(2) =0$ and $\rho(1) = 1$ we have
 \[\rho(g^\infty(1)) = \tl(\fib).\]
\end{theorem}
\begin{proof}
Recall $\fib = f^\infty(0)$ for $f(0) = 01, f(1) = 0$. The following claim is directly proved by induction on $n$
\begin{quote}
{\bf Claim:} If $f^n(0) = 0v_n$ and $f^n(1) = 0u_n$, then $\rho(g^n(1)) = v_n 0$ and $\rho(g^n(0)) = u_n 0$.
\end{quote}
Using this claim we obtain
\[ \tl(\fib) = 1 f(1) f^2(1) f^3(1) \cdots = 10 u_1 0 u_2 0 u_3 \cdots = \rho (1 0 g(0) g^2(0) \cdots) = \rho(g^\infty(1)).\]
\end{proof}

Due to the shape of the definition of $g$, the tree structure of this representation $\rho(g^\infty(1))$ of $\tl(\fib)$ is exactly the same as the tree structure of $\fib = f^\infty(0)$. Even more, using the same function $g$ and only another coding $\mu$ defined by $\mu(0) = 1$ and $\mu(1) = \mu(0) = 0$ yields $\mu(g^\infty(1)) = \fib$. So both $\fib$ and $\tl(\fib)$ can be obtained by applying two distinct codings on the same pure morphic sequence $g^\infty(1)$.

We elaborated this property only for $\fib$, but the construction of Theorem \ref{thmtlfib} can be generalized to arbitrary pure morphic sequences $\sigma$. More precisely, building the tree for $\tl(\sigma)$ using the tree structure of $\sigma$ yields a tree in which every subtree is shifted one position to the left compared to the tree for $\sigma$, and the remaining positions on the right can be filled in only finitely many ways bounded by the size of the alphabet, from which rationality of the tree/term follows. Working out further details and generalizations is a topic of further research.

\section{Conclusions, open problems}
\label{secconcl}

The main result of this paper is the series of alternative characterizations of morphic sequences: Theorems \ref{thmaut}, \ref{thmsubseq} and \ref{thmterm}. They all give if-and-only-if characterizations of morphic sequences, just like in any text book on automata theory you find a range of if-and-only-if characterizations of regular languages. The most remarkable theorem is Theorem \ref{thmaut}, characterizing morphic sequences by mix-DFAOs, exactly the same notion of automata defining mix-automatic sequences in a slightly different way.
In this paper we focus on the characterizations themselves. Applying them, like applying finiteness of DFAs gives rise to the pumping lemma as a standard technique to prove that a particular language is not regular, is a topic of further research.

The tree structure of a representation of a morphic sequence as described in Section \ref{sectree} can be seen as an unwinding of the mix-DFAO as used in Theorem \ref{thmaut}. Many distinct morphic sequences, like all 2-automatic sequences, may have representations with the same tree structure. This follows from the fact that for every $k$ a sequence is $k$-automatic if and only if it can be represented by the tree structure in which every node has degree $k$. Now Cobham's Theorem (\cite{AS03}, Theorem 11.2.2) implies that every not utimately periodic 2-automatic sequence does not have a representation with the tree structure in which every node has degree 3. A topic of further research is to investigate which tree structures may occur for some given morphic sequence. In Section \ref{secdistrepr} we already saw distinct tree representations for the sequence $\tl(\fib)$.

As a more complicated operation than $\tl$ (only removing the first element) we consider the operation $\even$ defined by
\[ \even(\sigma) \; = \; \sigma(0) \sigma(2) \sigma(4) \sigma(6) \cdots,\]
removing infinitely many elements. From \cite{AS03}, Theorem 7.9.1, it follows that every arithmetic subsequence of a morphic sequence, like $\even(\fib)$, is morphic too. So a natural question now is to explicitly represent $\even(\fib)$ as a morphic sequence in one of our formats. In a personal communication, Henk Don showed that $\even(\fib) = \tau(f^\infty(0))$ for $f, \tau$ defined by
\[ f(0) = 0122, f(1) = 01220, f(2) = 0120, \tau(0) = \tau(1) = 0, \tau(2) = 1.\]
This representation has a tree structure distinct from our tree structure of $\fib$. The following representation $\rho(g^\infty(0))$ for $g, \rho$ defined by
\[ g(0) = 01, g(1) = 2, g(2) = 31, g(3) = 04, g(4) = 0, \rho(0) = \rho(1) = 0, \rho(2) = \rho(3) = \rho(4) = 1\]
has the same tree structure as our tree structure of $\fib$. We are convinced that $\even(\fib) = \rho(g^\infty(0))$, but failed to give an elegant proof. We expect that it is beyond the power of the techniques from \cite{ZE11} to do this automatically.

A more general question is: given a morphic sequence $\sigma$ with some tree structure, can every arithmetic subsequence of $\sigma$ be represented by the same tree structure? A way to search for such a representation for explicit examples is by using Theorem \ref{thmterm}: build the tree of the given tree structure until some level, divide the elements of the intended arithmetic subsequence in a breadth-first manner, and check by a simple program which subtrees of this truncated part of the infinite tree match. This is the way how our representation $\rho(g^\infty(0))$ was found.

{\bf Acknowledgements:} We want to thank Henk Don and Wieb Bosma for fruitful discussions and for fruitful remarks on a preliminary version of this paper.

\bibliography{reftrs}

\end{document}